\documentclass{article}
\newfont{\msbm}{msbm10}
\newfont{\cmss}{cmss10}
\newtheorem{lem}{Lemma}
\newtheorem{theo}{Theorem}
\newtheorem{deff}{Definition}
\def\ba{\begin{eqnarray}}
\def\ea{\end{eqnarray}}
\def\be{\begin{equation}}
\def\ee{\end{equation}}

\def\L{{\cal L}}
\def\A{{\cal A}}
\def\S{{\cal S}}
\def\E{{\cal E}}
\def\H{{\cal H}}
\def\G{{\cal G}}
\def\hg{{\cal HG}}
\def\ag{{{\cal A}/{\cal G}}}
\def\agb{{\overline \ag}}
\def\Sig{\Sigma}

\def\Si{\mbox{$\Sigma$}}
\def\b{\beta}
\def\s{\sigma}
\def\gl{{\rm g}}
\def\sX{\hbox{\cmss X}}
\def\C{\hbox{\msbm C}}

\begin{document}
\title{Some properties of generalized connections \\ in quantum gravity}
\author{J. M. Velhinho}
\date{{\it Centra-UAlg}\\~\\
{\footnotesize {\it Present address:} Dep. de F\'\i sica,
Universidade da Beira Interior,\\
R. Marqu\^es d'\'Avila e Bolama,
6201-001 Covilh\~a, Portugal\\
E-mail: jvelhi@mercury.ubi.pt}}
\maketitle
\begin{abstract}
\noindent The quantum completion $\bar \A$ of the space of connections in a
manifold can be seen as the set of all morphisms from the 
groupoid of the edges of the manifold
to the (compact) gauge group. This algebraic construction  
generalizes the analogous description
of the gauge-invariant quantum configuration space $\agb$ 
of Ashtekar and Isham, 
clarifying the relation between the two spaces. Using this setup, 
we present a characterization of $\bar \A$
which brings the  gauge-invariant
degrees of freedom to the
foreground, thus making the action of the gauge group more transparent.
\end{abstract}
\section{Introduction}
\label{int}
Theories of connections play an important role in the description
of fundamental interactions, including Yang-Mills theories and 
gravity in the Ashtekar formulation~\cite{As,ALMMT}. 
Typically in such cases, the classical 
configuration space $\ag$ of connections modulo gauge transformations is
an infinite dimensional non-linear space of great complexity,
challenging the usual field quantization techniques.

Having in mind a rigorous quantization of theories of connections
and eventually of gravity,
methods of calculus in an extension
of $\ag$ have been developed over the last decade. 
For a compact gauge group $G$,
the extension $\agb$ is a natural compact
measurable space~\cite{AI}, allowing the construction of diffeomorphism
invariant measures~\cite{AL1,AL2,B2}.
An extension $\bar\A$ of the
space $\A$ of connections was also considered~\cite{B1}. 
In this case one still has to
divide by the appropriate action of gauge transformations. Besides being 
equally relevant for integral calculus, the space $\bar\A$ is particularly
useful for the definition of differential calculus in $\agb$, fundamental
in the construction of quantum observables~\cite{AL3}.

These developments rely crucially on the use of Wilson variables
(and generalizations), bringing to the foreground the important role of
parallel transport defined by certain types of curves. In this contribution
we will consider only the case of piecewise analytic curves, for which the
formalism was originally introduced, although most of the arguments apply
equally well to the more general piecewise smooth case~\cite{BS,LT}.
For both $\bar\A$ and $\agb$
one considers functions on $\A$ of the form
\be
\label{seccao10.1eqB}
\A\ni A\mapsto F\bigl(h(c_1,A),\ldots,h(c_n,A)\bigr)\,,
\ee
where $h(c,A)$ denotes the parallel transport defined by the curve $c$ and
$F:G^n\to\C$ is a continuous function. 
These functions define (overcomplete) coordinates on $\A$. In the case 
of $\agb$ only
closed curves -- loops -- are needed, producing
gauge invariant functions, or coordinates on $\ag$.
For compact $G$, the set of all functions (\ref{seccao10.1eqB})
is naturally a normed commutative $\ast$-algebra with identity. Therefore, 
the norm completion of this $\ast$-algebra
can be identified with  the $C^{\ast}$-algebra of 
continuous functions on a compact
space called the spectrum of the algebra.
This is precisely how the spaces $\agb$ and $\bar\A$ were originally
introduced: $\agb$ is the spectrum of the above algebra in the loop case,
and $\bar\A$ is likewise associated with more general open curves.
These spaces are natural completions of $\A/\G$ and $\A$, respectively,
and appear as
good candidates to replace them in the quantum context~\cite{AI,B1}.

It turns out that both $\agb$ and $\bar\A$ 
can also be seen as projective limits
of families of finite dimensional compact manifolds~\cite{AL1,AL2,MM}.
This projective characterization gives us a great deal of control
over the rather complex spaces $\agb$ and $\bar\A$, 
allowing the construction of
measures and vector fields on these spaces, starting from  corresponding
structures on the members of the projective families~\cite{AL1,AL2,B1,AL3,MM}.

For the case of $\agb$, a distinguished group of equivalence classes of loops,
called the hoop group $\hg$~\cite{AL1}, plays an important role:
$\agb$ can in fact be seen as the set of all homomorphisms $\hg\to G$, 
divided by the action of $G$~\cite{AL1}. As pointed out by Baez~\cite{B3}, 
for $\bar\A$ a similar
role is played by a certain groupoid. 
In our opinion however, 
this groupoid associated
to open curves has not yet occupied the place it deserves in the literature,
possibly due to the fact that groupoids have been introduced in the current 
mathematical physics literature only recently. Recall that a
groupoid is a category such that all arrows are invertible. 
Therefore, a groupoid
generalizes the notion of a group, in the sense that a binary operation with
inverse is defined, the difference being that not all pairs of elements can\
be composed.

In this contribution we consider the projective characterization of $\bar\A$ 
using the language of groupoids from the very beginning and show, 
using this 
formalism, that the quotient of $\bar\A$ by the action of the gauge group is 
homeomorphic to $\agb$. This new proof, establishing directly the equivalence 
at the projective limit level, is physically more transparent than the proof 
one can obtain by combination of results by Ashtekar and 
Lewandowski~\cite{AL1,AL2,AL3}, Marolf and Mour\~ao~\cite{MM} and 
Baez~\cite{B1}.
\section{Projective characterization of $\bar\A$}
\label{S1}
\subsection{Edge groupoid}
\label{seg}
Let \Si\ be an analytic, connected and orientable $d$-manifold. Let $\E$
denote the set of all continuous, oriented and piecewise analytic
parametrized curves in \Si , i.e.~continuous maps
$$
c:[0,t_1]\cup\ldots\cup [t_{n-1},1]\to\Sig
$$
such that the images $c\bigl(]t_k,t_{k+1}[
\bigr)$ are analytic submanifolds
embedded in \Si .  Let $\s:\E\to\Si$ be the map given by $\s(c)=c\bigl([0,1]
\bigr)$. The maps $s$ (source) and $r$ (range) are
defined, respectively, by $s(c)=c(0)$, $r(c)=c(1)$, $c\in\E$. Given two curves
$c_1,c_2\in\E$ such that $s(c_2)=r(c_1)$, let $c_2c_1$ denote 
the natural composition:
$$
(c_2c_1)(t)=\left\{\begin{array}{lll} c_1(2t), & {\rm for} & t\in[0,1/2] \\
c_2(2t-1), & {\rm for} & t\in[1/2,1]\,. \end{array} \right.
$$
Consider also the operation $c\mapsto c^{-1}$ 
given by $c^{-1}(t)=c(1-t)$. Strictly speaking, the composition
of parametrized curves is not associative, since the curves $(c_3c_2)c_1$ 
and $c_3(c_2c_1)$ are related by a reparametrization.
Also, the curve $c^{-1}$ is not the inverse of $c$.
Following Isham, Ashtekar and Lewandowski~\cite{AI,AL1} and Baez~\cite{B3},
we describe next an equivalence relation in $\E$ such that the corresponding
set of equivalence classes is a well defined groupoid~\cite{B3}, generalizing 
the hoop group~\cite{AL1}.

Let $G$ be a (finite dimensional) connected and compact Lie group and 
let $P(\Sig ,G)$ be a principal $G$-bundle over \Si . For simplicity we assume
that the bundle is trivial and that a fixed trivialization has been chosen.
Let $\A$ be the space of smooth connections on this bundle.
The parallel transport associated with a given connection $A\in\A$ and
a given curve $c\in\E$ will be denoted by $h(c,A)$.
\smallskip
\begin{deff}
\label{def1}
Two elements $c$ and $c'$ of $\E$ are said to be equivalent if
\begin{itemize}
\item[{\rm (}i\/{\rm )}] $s(c)=s(c'),\ \ r(c)=r(c')$
\item[{\rm (}ii\/{\rm )}] $h(c,A)=h(c',A)$, $\forall A\in\A$.
\end{itemize}
\end{deff}
Two curves related by  reparametrization are equivalent and the same is true
for curves $c$ and $c'$ that can be written in the form $c=c_2c_1$,
$c'=c_2c_3^{-1}c_3c_1$. For noncommutative G, 
these two conditions are equivalent to ({\it ii\/})~\cite{AL2,LT},
and therefore 
the equivalence relation is in fact the same for every
noncommutative compact Lie group. 
We will consider noncommutative groups from now on and denote the set of
all above defined equivalence classes by $\E\G$.
It is clear by ({\it i\/}) that the maps $s$ and $r$ are well
defined in $\E\G$. 
The map $\s$ can still be defined for special elements called edges. 
By edges we mean elements $e\in\E\G$ which are equivalence classes of
analytic (in all domain) curves $c:[0,1]\to\Si$. It is clear that the images
$c_1\bigl([0,1]\bigr)$ and $c_2\bigl([0,1]\bigr)$ of two equivalent analytic
curves coincide, and therefore we define $\s(e)$ as being $\s(c)$, where
$c$ is any analytic curve in the classe of the edge $e$.

We discuss next the natural groupoid structure on the set $\E\G$.
We will follow the terminology of category theory and refer to elements
of $\E\G$ as arrows. Given $\gamma,\gamma'\in\E\G$ 
such that $s(\gamma')=r(\gamma)$, one defines the composition $\gamma'\gamma$ 
by the composition of elements of $\E$. This operation is clearly well
defined and is now associative. The points of $\Si$ are called objects
in this context. Objects are in one-to-one correspondence with identity
arrows:  given $x\in\Sig$ the corresponding identity ${\bf 1}_x\in\E\G$ is the
equivalence class of $c^{-1}c$, with $c\in\E$ such that $s(c)=x$. 
If $\gamma$ is 
the class of $c$ then $\gamma^{-1}$ is the class of $c^{-1}$. It is clear that
$\gamma^{-1}\gamma={\bf 1}_{s(\gamma)}$ and 
$\gamma\gamma^{-1}={\bf 1}_{r(\gamma)}$. One therefore has a well defined 
groupoid, whose set of objects is $\Si$ and whose set of arrows is $\E\G$.
As usual, we will use the same notation -- $\E\G$ -- both for
the set of arrows and for the groupoid.
Notice that
every element $\gamma\in\E\G$ can be obtained as a composition of edges:
the groupoid 
$\E\G$ is generated by  edges, although
it is not freely generated.

For  $x,y\in\Si$, let ${\rm Hom}\,[x,y]$ be the set of all arrows $\gamma$
such that $s(\gamma)=x$ and $r(\gamma)=y$. 
It is clear that  ${\rm Hom}\,[x,x]$ 
is a group, $\forall x$. Since \Si\ is 
connected, the groupoid $\E\G$ is  connected, 
i.e.~${\rm Hom}\,[x,y]$ is a non-empty set $\forall x,y$. 
In this case,
any two groups ${\rm Hom}\,[x,x]$ and ${\rm Hom}\,[y,y]$ are
isomorphic.
Let us fix a point $x_0\in\Sig$ and consider the group ${\rm Hom}\,[x_0,x_0]$.
This group is precisely the so-called hoop group $\H\G$~\cite{AL1}, 
whose elements are equivalence classes 
of piecewise analytic loops. The elements of
${\rm Hom}\,[x_0,x_0]$ are called hoops and the identity 
arrow ${\bf 1}_{x_0}$ will be called the trivial hoop.

Given that $\E\G$ is connected, its elements may be written as
compositions of elements of ${\rm Hom}\,[x_0,x_0]$ and of an appropriate
subset of the set of all arrows:
\smallskip
\begin{lem}
\label{lem1}
Suppose that an unique arrow $\gamma_x\in{\rm Hom}\,[x_0,x]$ is given
for each $x\in\Sig$, $\gamma_{x_0}$ being the trivial hoop.
Then for every $\gamma\in\E\G$ there is a uniquely defined 
$\b\in{\rm Hom}\,[x_0,x_0]$ 
such that $\gamma=\gamma_{r(\gamma)}\b\gamma_{s(\gamma)}^{-1}$.
\end{lem}
\smallskip
This result can  be obviously adapted for any connected subgroupoid 
$\Gamma\subset\E\G$. The
converse of this result is the following lemma, where
${\rm Hom}_{\,\Gamma}\,[x_0,x_0]$ denotes the subgroup of the hoops that belong
to $\Gamma$.
\smallskip
\begin{lem}
\label{lem2}
Let $F$ be a subgroup of ${\rm Hom}\,[x_0,x_0]$ and $X\subset\Sig$ be a
subset of \Si\ such that $x_0\in X$. Suppose that an unique arrow
$\gamma_x\in{\rm Hom}\,[x_0,x]$ is given for each $x\in X$, 
$\gamma_{x_0}$ being the trivial hoop. Then the set $\Gamma$ of 
all arrows of the form
$\gamma_x\b\gamma_y^{-1}$, with $\b\in F$ and 
$x,y\in X$, is a connected subgroupoid of $\E\G$, and the group
${\rm Hom}_{\,\Gamma}\,[x_0,x_0]$ coincides with $F$.
\end{lem}
\subsection{$\bar \A$ as a projective limit}
\label{ssi}
By  condition ({\it ii\/}) in definition~\ref{def1},
the parallel transport is well defined for any element of $\E\G$. To
emphasize the algebraic role of connections and to simplify the notation,
we will denote by $A(\gamma)$ the parallel transport $h(c,A)$ defined
by $A\in\A$ and any curve $c$ in the  class $\gamma$. Since
the bundle $P(\Sig,G)$ is trivial, 
$A(\gamma)$ defines an element of the group $G$.
For every connection $A\in\A$, the map $\E\G \to G$ given by
\be
\label{grupg4}
\gamma \mapsto A(\gamma)
\ee
is a groupoid morphism, i.e.~$A(\gamma'\gamma)=A(\gamma')A(\gamma)$ and 
$A(\gamma ^{-1})=A(\gamma)^{-1}$. Thus, there is an  injective (but 
not surjective~\cite{AI,AL1,B1}) map from $\A$ to the set ${\rm Hom}\,
[\E\G,G]$ of all morphisms from $\E\G$ to $G$. 
It turns out that ${\rm Hom}\,[\E\G,G]$, 
when equipped with an appropriate topology,
is homeomorphic to the space $\bar\A$ of generalized 
connections~\cite{MM,AL2,B3}.
This identification can be proved using the fact that 
${\rm Hom}\,[\E\G,G]$ is the projective limit of a projective family
labeled by graphs in the 
manifold $\Si$~\cite{ALMMT,AL3}.
In what follows we will rephrase the projective characterization of 
${\rm Hom}\,[\E\G,G]$
using the language of groupoids. Later on we shall consider the action
of  local gauge transformations on ${\rm Hom}\,[\E\G,G]$ and  show,
in the context of projective methods, that the quotient of 
${\rm Hom}\,[\E\G,G]$ by this action is homeomorphic to $\agb$.

We start with the  set of labels for the projective family leading to
${\rm Hom}\,[\E\G,G]$, using
the notion of  independent edges~\cite{AL1}.
\smallskip
\begin{deff}
\label{def2}
A finite set $\{e_1,\ldots,e_n\}$ of edges is said to be independent if the
edges $e_i$ can intersect each other only at the points $s(e_i)$ or 
$r(e_i)$, $i=1,\ldots,n$.
\end{deff}
\smallskip
\noindent Let $\E\G\{e_1,\ldots,e_n\}$ be the subgroupoid 
of $\E\G$ freely
generated by the independent set $\{e_1,\ldots,e_n\}$, i.e.~the subgroupoid 
whose objects are all the
points $s(e_i)$ and $r(e_i)$, $i=1,\ldots,n$, and whose arrows are all possible
compositions of edges  $e_i$ and their inverses. 
Let $\L$ denote the set of all such subgroupoids.
Clearly, the sets $\{e_1,\ldots,e_n\}$ and
$\bigl\{e_1^{\epsilon_1},\ldots,e_n^{\epsilon_n}\bigr\}$, where
$e_j^{\epsilon_j}=e_j$ or $e_j^{-1}$, generate
the same subgroupoid.
Thus, a groupoid
$L\in\L$ is uniquely defined by a set $\bigl\{\s(e_1),\ldots,\s(e_n)\bigr\}$
of images of a set of independent edges. Notice that the union of the images 
$\s(e_i)$ is a graph in \Si , thus establishing the relation
with the more usual approach using graphs~\cite{B1,B2,AL3}.
The set $\L$ is a directed set, 
i.e.~for any given $L,L'\in \L$ there exists $L''\in\L$ such that both
$L$ and $L'$ are subgroupoids of $L''$, $L''\supset L$ and $ L''\supset L'$.
This follows from the crucial fact that for
every finitely generated subgroupoid $\Gamma\subset\E\G$ there is an
element $L\in\L$ such that $\Gamma\subset L$, which can be easily proved in the
piecewise analytic case~\cite{AL1}. 

Let us now consider the projective family. For each $L\in \L$,
let $\A_L:={\rm Hom}\,[L,G]$ be the set of all morphisms from the groupoid
$L$ to the group $G$. The family of spaces
$\A_L,\ L\in\L$, is a so-called compact Hausdorff projective 
family~\cite{AL2}, meaning that each of the spaces $\A_L$ is a compact
Hausdorff space and that  $\forall L,L'\in\L$ such that $L'\supset L$ there exists a
surjective and continuous projection $p_{L,L'}:\A_{L'}\rightarrow \A_L$
satisfying
\be
\label{grupg6}
p_{L,L''}=p_{L,L'}\circ p_{L',L''},\ \forall L''\supset L'\supset L\,.
\ee
There is a well defined notion of limit of the family of spaces
$\A_L$ -- the  projective limit -- which is also a compact Hausdorff
space. We discuss next some aspects of this construction.

Let $L=\E\G\{e_1,\ldots,e_n\}$ be an element of $\L$. 
Since the morphisms $L\to G$ are
determined by the images of the generators, one gets a
bijection $\rho_{e_1,\ldots,e_n}:\A_L\to G^n$, 
\be
\label{bara4}
\A_L\ni\bar A\mapsto \bigl(\bar A(e_1),\ldots,\bar A(e_n)\bigr)\in G^n\,.
\ee
Through this identification with $G^n$, the space $\A_L$ can be seen
as a compact Hausdorff space.  For 
$L'\supset L$ the projection 
$p_{L,L'}:\A_{L'}\to \A_L$ is defined as the map
sending each element of $\A_{L'}$ to its restriction to $L$.  
Using the maps (\ref{bara4}) it is 
not difficult to see that the projections $p_{L,L'}$ are surjective and
continuous~\cite{AL2,V}.

The projective limit of the family $\{\A_L,p_{L,L'}\}$ is the
subset $\A_{\infty}$ of the cartesian product \sX$_{L\in\L}\A_L$ of the
elements $(A_L)_{L\in\L}$ satisfying the consistency conditions:
\be
\label{eq1}
p_{L,L'}A_{L'}=A_L\,,\ \ \ \forall \, L'\supset L\,.
\ee
The cartesian product is a compact Hausdorff space with respect to the 
Tychonov product
topology. Given the continuity of the projections $p_{L,L'}$, the projective 
limit $\A_{\infty}$ is a closed subset~\cite{MM,AL2} and therefore is also
compact Hausdorff. The induced topology in $\A_{\infty}$
is the weakest topology such that all the following projections are continuous:
\ba
p_L:\qquad\quad \A_{\infty} & \to & \A_L \nonumber \\
\label{eqa1}
(A_L)_{L\in\L} & \mapsto & A_L\,.
\ea
The proof that the projective limit $\A_{\infty}$ coincides with the set of 
all groupoid morphisms ${\rm Hom}\,[\E\G,G]$ follows essentially the same
steps as  the proof of the well known fact that the algebraic dual
of any vector space is a projective limit.
In what follows we will identify
$\A_{\infty}$ with ${\rm Hom}\,[\E\G,G]$. For simplicity,
we will refer to the induced topology on ${\rm Hom}\,[\E\G,G]$ as the
Tychonov topology.
\section{Equivalence of the projective characterizations of $\bar\A/\bar \G$ 
and $\agb$}
\label{sgt}
In this section we will study the relation between the space of
generalized connections considered above and the space $\agb$ of
generalized connections modulo gauge transformations,
{}from the point of view of projective techniques. The gauge
transformations act naturally on ${\rm Hom}\,[\E\G,G]$ and, as expected,
the quotient of ${\rm Hom}\,[\E\G,G]$ by this action is homeomorphic to
$\agb$. The proof presented here complements previous 
results~\cite{AL1,AL2,MM,B1,AL3} and clarifies the relation between the
two spaces. The introduction of the groupoid $\E\G$ plays a relevant role in 
this result.

We start with a brief review of the projective characterization of 
$\agb$~\cite{AL1,AL2,MM}. In this case the projective family
is labeled by certain ``tame'' subgroups of the hoop group $\H\G\equiv
{\rm Hom}\,[x_0,x_0]$, which are subgroups freely generated by finite sets of 
independent hoops. We will 
denote the family of such subgroups by $\S_{\H}$. For each $S\in\S_{\H}$
one considers the set $\chi_S$ of all homomorphisms $S\to G$,
$\chi_S:={\rm Hom}[S,G]$.
The family $\{\chi_S\}_{S\in\S_{\H}}$ is a compact Hausdorff
projective family, whose projective limit is ${\rm Hom}\,[\H\G,G]$,
the set of all homomorphisms $\H\G\to G$~\cite{AL2}.
The space ${\rm Hom}\,[\H\G,G]$
is  equiped with a Tychonov-like topology, namely the
weakest topology such that all the natural projections
\be
\label{hoopint4}
p_S:{\rm Hom}[\H\G,G]\to\chi_S\,,\ \ S\in\S_{\H}\,,
\ee
are continuous.
The group $G$ acts continuously on ${\rm Hom}\,[\H\G,G]$~\cite{AL2}: 
\be
\label{hoopint6}
{\rm Hom}\,[\H\G,G]\times G\ni(H,g)\mapsto H_g\,:\ H_g(\b)=g^{-1}
H(\b)g,\ \forall\b\in\H\G.
\ee
This action corresponds
to the non-trivial part of the action of the group of generalized local
gauge transformations (see below). 
It is a well established fact that the quotient space 
${\rm Hom}\,[\H\G,G]/G$ is homeomorphic to $\agb$~\cite{MM,AL2}. 

Let us consider now the corresponding action of local gauge transformations
on generalized connections. The group of local gauge transformations
associated with the structure group $G$ is the group $\G$ of all smooth
maps $\gl:\Sig\to G$, acting on smooth connections as follows:
$$
\A\ni A\mapsto\gl^{-1}A\gl + \gl^{-1}d\gl\,,
$$
where $d$ denotes the exterior derivative. The corresponding action on
parallel transports $A(\gamma)$ defined by $A\in\A$ and $\gamma\in\E\G$
is given by
\be
\label{hoopint7}
A(\gamma)\mapsto \gl(x_2)^{-1}A(\gamma)\gl(x_1)\, \ \gl\in\G\, ,
\ee
where $x_1=s(\gamma)$, $x_2=r(\gamma)$. Let us consider the extension
$\bar\G$ of $\G$,
\be
\label{hoopint8a}
\bar\G={\rm Map}[\Sigma,G]=G^{\Sigma}\cong\sX_{x\in\Sigma}G_x\,,
\ee
of all maps $\gl:\Sig\to G$, not necessarily smooth or even continuous. 
This group $\bar\G$ of ``generalized local gauge transformations'' acts
naturally on the space of generalized connections ${\rm Hom}\,[\E\G,G]$,
\be
\label{hoopint7a}
{\rm Hom}\,[\E\G,G]\times \bar \G\ni(\bar A,\gl)\mapsto\bar A_{\gl}\in
{\rm Hom}\,[\E\G,G]
\ee
where
\be
\label{hoopint8}
\bar A_{\gl}(\gamma)=
\gl\bigl(r(\gamma)\bigr)^{-1}\bar A(\gamma)\gl\bigl(s(\gamma)\bigr),\ 
\forall\gamma\in\E\G\,,
\ee
generalizing (\ref{hoopint7}). It is natural to consider the quotient of
${\rm Hom}[\E\G,G]$ by the action of $\bar\G$, since ${\rm Hom}[\E\G,G]$
is also made of all the morphisms $\E\G\to G$, without any continuity 
condition. The group $\bar\G$ is compact Hausdorff  and its action 
is continuous~\cite{AL2,AL3}. Therefore
${\rm Hom}[\E\G,G]/\bar\G$ is also a compact Hausdorff space.

Let us consider  the compact space $\bar\A$ introduced by Baez as
the spectrum of a $C^*$-algebra,
obtained by completion of a $*$-algebra of functions in $\A$~\cite{B1}. 
The original $C^*$-algebra can then be identified with the
algebra $C(\bar\A)$ of continuous functions in $\bar\A$. The group of 
local gauge transformations
acts naturally on $C(\bar\A)$ and the subspace 
$C^{\G}(\bar\A)\subset C(\bar\A)$ 
of gauge invariant functions is also a commutative $C^*$-algebra
with identity~\cite{B1},
whose spectrum we will denote by $\bar\A/\bar\G$.

One therefore has four extensions of the classical configuration space 
$\A/\G$, namely $\agb$, $\bar\A/\bar\G$, ${\rm Hom}\,[\H\G,G]/G$ and
${\rm Hom}\,[\E\G,G]/{\bar\G}$. The first two spaces are tied to the
$C^*$-algebra formalism whereas the last two appear in the context of 
projective methods. As expected, all these spaces are naturally homeomorphic.
Let us consider the following diagram
\smallskip
$$
\begin{array}{ccc}
\agb & \longleftrightarrow & {\rm Hom}\,[\H\G,G]/G \\
& & \\
\updownarrow & & \\
& & \\
\bar\A/\bar\G & \longleftrightarrow & {\rm Hom}\,[\E\G,G]/\bar\G
\end{array}
$$

\smallskip
\noindent The correspondence between $\agb$ and ${\rm Hom}\,[\H\G,G]/G$ was
established by Marolf and Mour\~ao~\cite{MM}. A generalization of 
this  result produces a homeomorphism \-between $\bar\A$ and 
${\rm Hom}\,[\E\G,G]$~\cite{AL2}, which is easily seen to be
equivariant, leading to a homeomorphism between $\bar\A/\bar\G$ and 
$\,{\rm Hom}\,[\E\G,G]/\bar\G$~\cite{AL3}. The
correspondence between $\agb$ and $\bar\A/\bar\G$ follows from results
by Baez~\cite{B1}.
In the remaining of this  section we will show directly that 
${\rm Hom}\,[\E\G,G]/\bar\G$ 
is homeomorphic to ${\rm Hom}[\H\G,G]/G$. The relevance of this 
new proof of a known result lies in 
the clear relation established between ${\rm Hom}\,[\E\G,G]\ (\cong
\bar\A)$ and ${\rm Hom}\,[\H\G,G]/G\ (\cong\agb)$, without having to rely on
the characterization of these spaces as spectra of $C^*$-algebras. 

Since $\H\G\equiv{\rm Hom}\,[x_0,x_0]$ is a subgroup of the groupoid $\E\G$,
a projection from ${\rm Hom}\,[\E\G,G]$ to
${\rm Hom}\,[\H\G,G]$, given by the restriction of elements of 
${\rm Hom}\,[\E\G,G]$ to the group $\H\G$,
is naturally defined.
We will show that this projection
is surjective and equivariant with respect to the actions of $\bar\G$ on 
${\rm Hom}\,[\E\G,G]$ and ${\rm Hom}\,[\H\G,G]$, thus defining a map
${\rm Hom}\,[\E\G,G]/\bar\G\to{\rm Hom}\,[\H\G,G]/G$ 
which is in fact a bijection.
We will also present the more relevant elements of the proof 
that the latter map 
and its inverse are continuous. For this we will need some lemmas, 
whose proofs will appear elsewhere~\cite{V}.

We will start by showing that ${\rm Hom}\,[\E\G,G]$ is homeomorphic to
${\rm Hom}\,[\H\G,G]\times\bar\G_{x_0}$, where $\bar\G_{x_0}$ is the
subgroup of $\bar\G$ (\ref{hoopint8a}) of the elements $\gl$ such that 
$\gl(x_0)={\bf 1}$.
Let us fix  a unique edge $e_x\in{\rm Hom}\,[x_0,x]$ for
each $x\in\Sig$, $e_{x_0}$ being the trivial hoop. Let us denote
this set of edges by $\Lambda=\{e_x,\ x\in\Sig\}$. Consider the map
\be
\label{novo1}
\Theta_{\Lambda}:{\rm Hom}\,[\E\G,G]\to{\rm Hom}[\H\G,G]\times\bar\G_{x_0}
\ee
where $\bar A\in{\rm Hom}\,[\E\G,G]$ is mapped to $(H,\gl)\in{\rm Hom}\,
[\H\G,G]\times\bar\G_{x_0}$ such that
\be
\label{novo2}
H(\b)=\bar A(\b),\ \ \forall\b\in\H\G
\ee
and
\be
\label{novo3}
\gl(x)=\bar A\bigl(e_x\bigr),\ \ \forall x\in\Sigma\,.
\ee
Consider also the natural action of $\bar\G$ on ${\rm Hom}\,[\H\G,G]\times
\bar\G_{x_0}$ given by
\be
\label{hoopint50}
\bigl({\rm Hom}\,[\H\G,G]\times\bar\G_{x_0}\bigr)\times\bar\G\ni\bigl(
(H,\gl),\gl'\bigr)\mapsto (H_{\gl'},\gl_{\gl'})\,,
\ee
where
\be
\label{hoopint51}
H_{\gl'}(\b)=\gl'(x_0)^{-1}H(\b)\gl'(x_0),\ \ \forall\b\in
\H\G
\ee
and
\be
\label{hoopint52}
\gl_{\gl'}(x)=\gl'(x)^{-1}\gl(x)\gl'(x_0),\ \ \forall x\in\Sigma \,.
\ee
\begin{theo}
\label{teonovo1}
For any choice of the set $\Lambda$, the map $\Theta_{\Lambda}$ is a
homeomorphism, equivariant with respect to the action of $\bar\G$.
\end{theo}
\smallskip
It is fairly easy to see that $\Theta_{\Lambda}$ is bijective and
equivariant: for a given $\Lambda$, the map $\Theta_{\Lambda}$ is
clearly well defined and its inverse is given by
$(H,\gl)\mapsto\bar A$ where
\be
\label{novo4}
{\bar A}(\gamma)=\gl\bigl(r(\gamma)\bigr)H\Bigl(
e_{r(\gamma)}^{-1}\gamma\,e_{s(\gamma)}
\Bigr)\gl\bigl(s(\gamma)\bigr)^{-1},\ \ \forall\gamma\in\E\G\,.
\ee
It is  clear that $\Theta_{\Lambda}$ is equivariant with respect to
the action
of $\bar\G$ on ${\rm Hom}\,[\H\G,G]\times\bar\G_{x_0}$
(\ref{hoopint51}, 
\ref{hoopint52}) and on ${\rm Hom}\,[\E\G,G]$ (\ref{hoopint7a}, 
\ref{hoopint8}). It remains to be
shown that $\Theta_{\Lambda}$ is a homeomorphism. Recall that the
topologies of ${\rm Hom}\,[\H\G,G]$ and ${\rm Hom}\,[\E\G,G]$ are defined by
the projective families $\{\chi_S\}_{S\in\S_{\H}}$ and $\{\A_L\}_{L\in\L}$
considered previously.

Given $S\in\S_{\H}$ and $x\in\Sigma$, let $P_S$ and $\pi_x$, respectively, be 
the projections from ${\rm Hom}\,[\H\G,G]\times\bar\G_{x_0}$ to $\chi_S$ and
$G_x$ (the copy of $G$ associated with the point $x$). Recall that the
topology of ${\rm Hom}\,[\H\G,G]\times\bar\G_{x_0}$ is the weakest
topology such that all the maps $P_S$ and $\pi_x$ are continuous. So, $\Theta_
{\Lambda}$ is continuous if and only if the maps $P_S\circ\Theta_{\Lambda}$
and $\pi_x\circ\Theta_{\Lambda}$ are continuous, $\forall S\in\S_{\H}$ and
$\forall x\in\Sigma$. Likewise, $\Theta_{\Lambda}^{-1}$ is continuous if and 
only if all the maps $p_L\circ\Theta_{\Lambda}^{-1}:{\rm Hom}\,[\H\G,G]
\times\bar\G_{x_0}\to\A_L$ are continuous, where the projections 
$p_L:{\rm Hom}\,[\E\G,G]\to\A_L$ are defined in (\ref{eqa1}).

It is straightforward to show that the maps $\pi_x\circ\Theta_{\Lambda}$
are continuous: given $x\in\Sigma$, one just has to consider the
subgroupoid $L=\E\G\bigl\{e_x\bigr\}$ generated by the edge 
$e_x\in\Lambda$ and the homeomorphism (\ref{bara4}) 
$\rho_{e_x}:\A_L\to G$. It is clear that $\pi_x\circ\Theta_{\Lambda}$ 
coincides with $\rho_{e_x}\circ p_L$, being therefore continuous.

On the other hand, to show that $P_S\circ\Theta_{\Lambda}$ and 
$p_L\circ\Theta_{\Lambda}^{-1}$ are continuous one needs to consider 
explicitly the relation between the spaces $\A_L$ and 
$\chi_S$, $L\in\L$, $S\in\S_{\H}$~\cite{V}.
\smallskip
\begin{lem}
\label{prophoopint1}
For every $S\in\S_{\H}$ there exists a connected subgroupoid 
$L\in\L$ such that $S$ is a subgroup of $L$. The projection 
\be
\label{eqprop1}
p_{S,L}:\A_L\to\chi_S 
\ee
defined by the restriction of elements of $\A_L$ to  $S$ is 
continuous and satisfies
\be
\label{eqprop2}
P_S\circ\Theta_{\Lambda}=p_{S,L}\circ p_L\,  ,\ \ \forall \Lambda.
\ee
\end{lem}
The continuity of the maps $P_S\circ\Theta_{\Lambda}$, $\forall S\in
\S_{\H}$, follows immediately from lemma \ref{prophoopint1}. To show that 
the maps $p_L\circ\Theta_{\Lambda}^{-1}$ are continuous one needs the 
converse of lemma \ref{prophoopint1}. Recall that for a given  
subgroupoid $\Gamma\subset\E\G$, 
${\rm Obj}\,\Gamma$ denotes the set of  objects of $\Gamma$ (the set of all
points of \Si\ which are
range or source for some arrow in $\Gamma$) and that 
${\rm Hom}\,[\Gamma,G]$
stands for the set of all morphisms $\Gamma\to G$.  
We will also denote by $\Pi_{\Gamma}$ 
the natural projection
from $\bar\G_{x_0}$ to the subgroup $\bar\G_{x_0}(\Gamma)$ of all maps 
${\rm Obj}\,\Gamma\to G$ such that $\gl(x_0)={\bf 1}$.
\smallskip
\begin{lem}
\label{prophoopint2}
For every $L\in\L$ there exists $S\in\S_{\H}$ and a connected subgroupoid 
$\Gamma\subset\E\G$, with ${\rm Obj}\,\Gamma={\rm Obj}\,L\stackrel{.}{\cup}
\{x_0\}$, such that $L\subset\Gamma$ and ${\rm Hom}_{\,\Gamma}\,
[x_0,x_0]=S$. The natural projection from ${\rm Hom}\,[\Gamma,G]$ to $\A_L$ 
defines a map 
\be
\label{eqprop3}
p_{L,S}:\chi_S\times\bar\G_{x_0}(\Gamma)\to\A_L
\ee
which is continuous and such that for an appropriate choice of 
$\Lambda$ one has
\be
\label{eqprop4}
p_L\circ\Theta_{\Lambda}^{-1}=p_{L,S}\circ(p_S\times\Pi_{\Gamma})\,.
\ee
\end{lem}
Given that the projections $p_S:{\rm Hom}\,[\H\G,G]\to\chi_S$ and 
$\Pi_{\Gamma}:\bar\G_{x_0}\to\bar\G_{x_0}(\Gamma)$ are continuous, lemma
\ref{prophoopint2} shows that for every fixed $L\in\L$ there exists a  
${\Lambda}$ such that $p_L\circ\Theta_{\Lambda}^{-1}$ is continuous, which 
still does not prove that all the maps $p_L\circ\Theta_{\Lambda}^{-1}$
are continuous for a given ${\Lambda}$. Notice however that
the map $\Theta_{\Lambda}\circ\Theta_{\Lambda'}^{-1}$ is a homeomorphism
for any $\Lambda$ and $\Lambda'$~\cite{V}, 
from what
follows immediately that
\smallskip
\begin{lem}
\label{corhoopint2}
The continuity of $p_L\circ\Theta_{\Lambda}^{-1}$ is equivalent to the
continuity of $p_L\circ\Theta_{\Lambda'}^{-1}$, for any other values of 
$\Lambda'$.
\end{lem}
\smallskip
\noindent This lemma, together with lemma \ref{prophoopint2}, shows that, 
for a 
given $\Lambda$, all the maps $p_L\circ\Theta_{\Lambda}^{-1}$, $L\in\L$, 
are continuous, which concludes the proof of theorem \ref{teonovo1}.

The identification of ${\rm Hom}\,[\E\G,G]/\bar\G$ with ${\rm Hom}\,[\H\G,G]/G$
now follows easily. Consider a fixed $\Lambda$. Since $\Theta_{\Lambda}$
is a homeomorphism equivariant with respect to the continuous action of 
$\bar\G$, we conclude that
${\rm Hom}\,[\E\G,G]/\bar\G$ is homeomorphic to
$\bigl({\rm Hom}\,[\H\G,G]\times\bar\G_{x_0}\bigr)/\bar\G$. On the other
hand it is clear that
\begin{eqnarray}
\bigl({\rm Hom}\,[\H\G,G]\times\bar\G_{x_0}\bigr)/\bar\G & = &
\bigl({\rm Hom}\,[\H\G,G]/G\bigr)\times\bigl(\bar\G_{x_0}/\bar\G_{x_0}\bigr)
\cong \nonumber \\
\label{numsegunda}
&\cong & {\rm Hom}\,[\H\G,G]/G\,.
\end{eqnarray}
Thus, as a corollary of theorem \ref{teonovo1} one gets that
\smallskip
\begin{theo}
\label{corhoopint1}
The spaces ${\rm Hom}\,[\E\G,G]/\bar\G$ and ${\rm Hom}\,[\H\G,G]/G$ are
homeomorphic.
\end{theo}

\section*{Acknowledgements}
I would like to thank Jos\'e Mour\~ao, Paulo S\'a and Thomas Thiemann, for
encouragement and helpful discussions. This work was supported in part by
CENTRA/UAlg, PRAXIS 2/2.1/FIS/286/94 and CERN/P/FIS/15196/1999.


\begin{thebibliography}{99}

\bibitem{As} A. Ashtekar, {\it Lectures on non-Perturbative Canonical
Quantum Gravity} (World Scientific, Singapore, 1991).


\bibitem{ALMMT} A. Ashtekar, J. Lewandowski, D. Marolf,
J. Mour\~ao, T. Thiemann, {\em J. Math. Phys.} {\bf 36}, 6456 (1995).

\bibitem{AI} A. Ashtekar, C. J. Isham,
{\em Class. Quant. Grav.} {\bf 9}, 1433 (1992).

\bibitem{AL1} A. Ashtekar, J. Lewandowski, ``Representation
Theory of Analytic Holonomy $C^\star$ Algebras'' in {\em Knots and 
Quantum Gravity}, ed.\ J. Baez (Oxford University Press,
Oxford, 1994).


\bibitem{AL2} A. Ashtekar, J. Lewandowski, 
{\em J. Math. Phys.} {\bf 36}, 2170 (1995).

\bibitem{B2} J. Baez, ``Diffeomorphism Invariant Generalized
Measures on the Space of Connections Modulo Gauge Transformations'',
in {\em Proceedings of the Conference on Quantum
Topology}, ed. D. Yetter (World Scientific, Singapore, 1994). 


\bibitem{B1} J. Baez, {\em Lett. Math. Phys.} {\bf 31}, 213 (1994).

\bibitem{AL3} A. Ashtekar, J. Lewandowski, 
{\em J. Geom. Phys.} {\bf 17}, 191 (1995).

\bibitem{BS} J. Baez, S. Sawin, {\em J. Funct. Analysis} {\bf 158}, 253 (1998).

\bibitem{LT} J. Lewandowski, T. Thiemann, {\em Class. Quant. Grav.} {\bf 16}, 
2299 (1999).

\bibitem{MM}  D. Marolf, J. Mour\~ao, {\em Commun. Math. Phys.} 
{\bf 170}, 583 (1995).

\bibitem{B3} J. Baez, {\em Adv. Math.} {\bf 117}, 253 (1996).

\bibitem{V} J. M. Velhinho, ``A groupoid approach to spaces of 
generalized connections'', preprint arXiv: hep-th/0011200.

\end{thebibliography}
\end{document}